\documentclass[preprint,5p]{elsarticle}

\usepackage{bm}
\usepackage{amsmath}
\usepackage{graphicx}
\usepackage{booktabs}
\usepackage{url}
\usepackage[multiple]{footmisc}

\usepackage{listings}
\newcommand\fitsmap{\textit{FitsMap}}
\newcommand\mg{\textit{Map Generator}}
\newcommand\mv{\textit{Map Viewer}}

\journal{Astronomy And Computing}

\newcommand\aj{\ref@jnl{AJ}}
\newcommand\aap{\ref@jnl{A\&A}}

\begin{document}

\begin{frontmatter}

\title{FitsMap: A Simple, Lightweight Tool For Displaying Interactive Astronomical Image and Catalog Data}

\author[1]{Ryan Hausen\corref{mycorrespondingauthor}}
\cortext[mycorrespondingauthor]{Corresponding author}
\ead{rhausen@ucsc.edu}

\affiliation[1]{organization={Department of Computer Science and Engineering, University of California, Santa Cruz},
            addressline={1156 High Street},
            city={Santa Cruz},
            postcode={95064},
            state={CA},
            country={USA},
}

\author[2]{Brant E. Robertson}
\affiliation[2]{organization={Department of Astronomy and Astrophysics, University of California, Santa Cruz},
            addressline={1156 High Street},
            city={Santa Cruz},
            postcode={95064},
            state={CA},
            country={USA}
}

\begin{abstract}

The visual inspection of image and catalog data continues to be a valuable aspect
of astronomical data analysis. As the scale of astronomical image and catalog data
continues to grow, visualizing the data becomes increasingly difficult. In this
work, we introduce \fitsmap{}, a simple, lightweight tool for visualizing
astronomical image and catalog data. \fitsmap{} only requires a simple web
server and can scale to over gigapixel images with tens of millions of sources.
Further, the web-based visualizations can be viewed performantly on mobile
devices. \fitsmap{} is implemented in Python and is open source
(https://github.com/ryanhausen/fitsmap).

\end{abstract}




\begin{keyword}
    Astronomy web services, 
    Astronomy data visualization, 
    Astronomy data analysis, 
    Human-centered computing~Scientific visualization, 
    Human-centered computing~Visualization toolkits 
\end{keyword}

\end{frontmatter}

\section{Introduction}
\label{sec:introduction}

Astronomical image data is inherently visual, and visual inspection and
interpretation remain vital tools in the scientific process in astronomy.
Upcoming telescopes like the James Webb Space Telescope \citep[JWST; for a review, see][]{robertson2022a},
Nancy Grace Roman Space Telescope \citep{spergel2015a,akeson2019a}, and Vera
Rubin Observatory \citep{ivezic2008a,ivezic2019a} will produce larger
and deeper images of space than ever before. Specialized tools for visualizing
and interacting with large scale astronomical images will enable a more rapid
transition from observation to analysis for these facilities.

Visualizing large astronomical images is not a new problem, and several tools
have been developed to meet this need. There are desktop tools like \texttt{DS9}
\citep{joye2019} and \texttt{Aladin} \citep{bonnarel1994}, which are \textit{thick}
clients that render local images or fetch images from a remote server. The term
\textit{thick client} refers to programs that have little to no reliance on a
remote server to perform the computational
tasks associated with the software. Advances in web
technologies have also enabled the development of \textit{thin}
clients for astronomical image data visualization. \textit{Thin} clients are
dependent on a separate server to perform their computational tasks.
An advantage to using
\text{thin} clients is that by offloading heavier computational tasks to a
remote server the client hardware requirements are far less powerful than
the server requirements and can even include mobile devices.
Some examples of thin clients
developed for astronomical image and catalog visualization include
\texttt{VisiOmatic}
\citep{bertin2015,bertin2019}, \texttt{Aladin Lite} \citep{boch2014}, \texttt{RCSED}
\citep{Chilingarian2017, klochkov2021}, the GAIA archive visualization service
\citep{moitinho2017}, \texttt{Toyz} \citep{moolekamp2015},
WorldWide Telescope
\citep{rosenfield2018},
\texttt{ASTRODEEP} \citep{derriere2017,wassong2019}, and \texttt{ESASky}
\citep{giordano2018}. More recently, the popularity of the Python
\citep{rossum1995} programming language has inspired a series of new
\textit{thick} and \textit{thin} clients that leverage Python and the Jupyter
Lab/Notebook (iPython; Kluyver2016) ecosystem. Some examples include
\texttt{Vizic} \citep{yu2017},
\texttt{LSSGalPy} \citep{argudo2017}, Astro Data Lab
\citep{juneau2021},
\texttt{Jovial} \citep{araya2018}, and
\texttt{Jdaviz} \citep{osteen2021}.

\begin{sloppypar}
    This work introduces a new tool called \fitsmap{}. \fitsmap{} is a \text{thin}
    client that improves on previous methods by only requiring a simple web
    server for offloading computational tasks.
    \fitsmap{} is designed to work for a simple use case where the user has images
    (\texttt{FITS} \citep{wells1981}, \texttt{PNG} \citep{png2021},
    or \texttt{JPEG} \citep{jpg2019}), catalogs
    associated with the image data, and would like to view and possibly
    share that data interactively.
    A user can generate a website that displays image and catalog
    information using a single \fitsmap{} function. \fitsmap{} has already
    been deployed to visualize large simulated JWST \citep{williams2018a} and
    Roman \citep{drakos2021} datasets and machine-learning galaxy morphological
    analyses of large Hubble Space Telescope
    surveys \citep{hausen2020}.
\end{sloppypar}

The remainder of this paper is structured as follows. Section \ref{sec:methods}
describes the details of the design and methods of \fitsmap{}. Section
\ref{sec:performance} describes the performance of \fitsmap{} and, Section
\ref{sec:conclusion} reviews the contributions and future directions of
\fitsmap{}.

\section{Methods}
\label{sec:methods}

The design philosophy behind \fitsmap{} is to render visualizing image and catalog
data simple by minimizing the number of steps and technical knowledge required
to go from image and catalog files to an interactive display. The only
requirements to generate and view the output from \fitsmap{} are \fitsmap{}
itself and a
web server.
If viewing the output locally, \fitsmap{} can also run the web
server.
If \fitsmap{} is used to view image data alone a web server is not required,
as the user can simply open the output
\texttt{index.html} file using a browser and the processed images will render.

Catalog data present a unique challenge in the effort to minimize server
requirements. Small catalogs can be transferred from the server to
be clustered and
rendered on the client, but large catalogs practically cannot.
Large catalogs may either take too long to send usefully
or exceed the memory capabilities of the client. Managing the
catalog data on the server-side requires a more sophisticated web server that
clusters and serves catalog data as needed. Leveraging the continuing decline
in the cost of storage \citep{mccallum2022}, \fitsmap{} separates the catalog
data into a set of tiles at different zoom levels. \fitsmap{} then
precomputes clusters of the catalog data at
every zoom level for every tile (Section \ref{sec:mg-catalog-parsing}).
Precomputing the clustering and tiling the data significantly reduce the
computational requirements of both the server and client by never requiring the entire
catalog to be loaded into memory.

The \fitsmap{} architecture consists of two main components: the \mg{} and the
\mv{}. The \mg{}, implemented in Python, builds the website, including the \mv{}
files. The \mv{}, implemented in Javascript, renders the webpage and fetches the
image and catalog data.

\subsection{\mg{}}
\label{sec:mg}

\begin{figure*}[h]
    \centering
    \includegraphics[width=\textwidth]{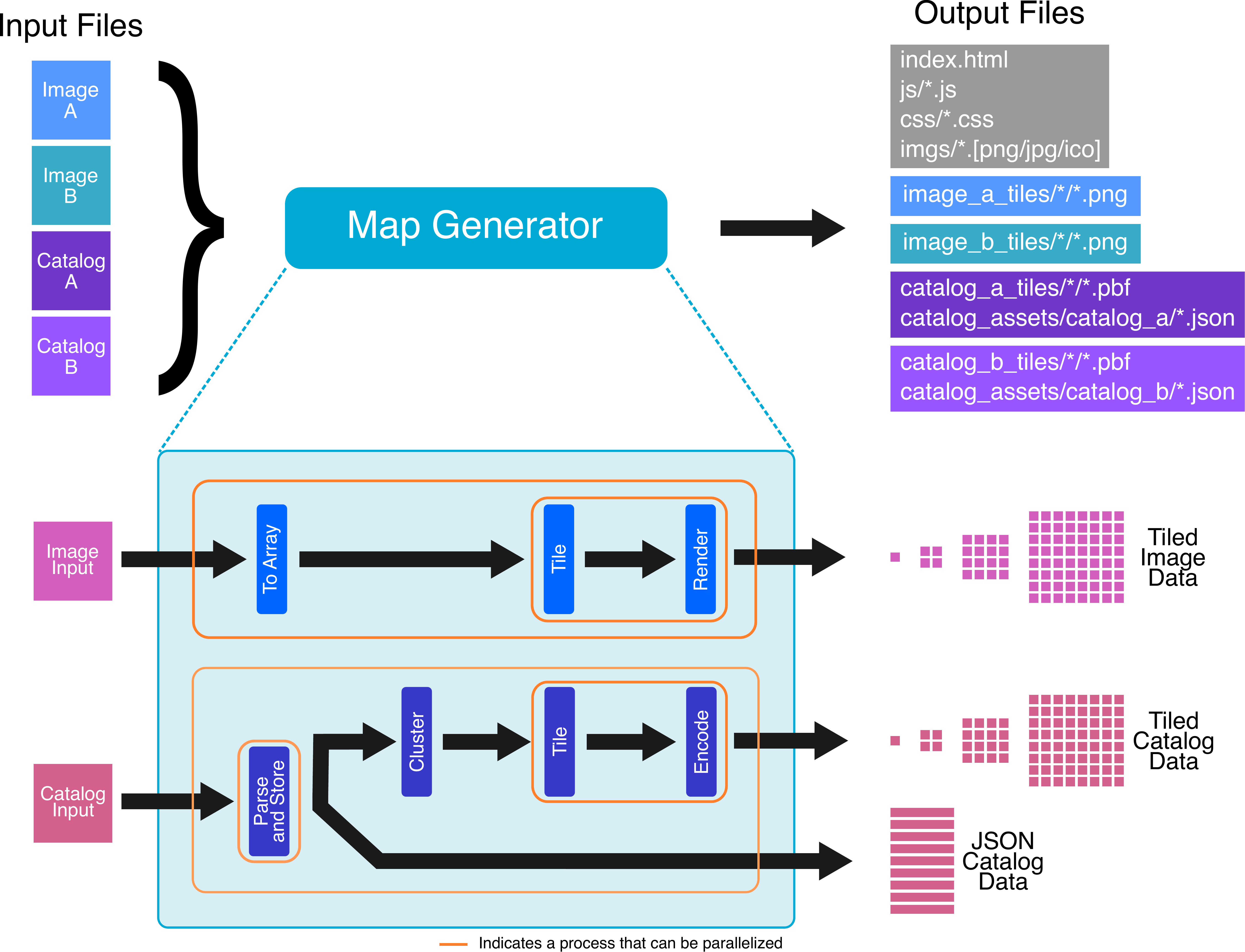}
    \caption{An overview of the \mg{} in the \fitsmap{} architecture. The \mg{},
             described in Section \ref{sec:mg}, processes the input image and
             catalog files and converts them into a format that can be rendered
             by the \mv{} (Section \ref{sec:mv}). The \mg{} can process multiple
             files in parallel and can further leverage parallelism when
             processing each file. The \mg{} builds a directory of files that
             contains the tiled image and catalog data along with the \mv{} data
             and code (gray). The output website can be viewed using a simple web
             server.}
    \label{fig:overview}
\end{figure*}

The \mg{} parses a list of input image and catalog file locations and converts
them into a format that the \mv{} can render. The output of the \mg{} is a
directory that contains the processed input image and catalog data and the
supporting HyperText Markup Language (\texttt{HTML}), images, Cascading Style Sheets
(\texttt{CSS}), and JavaScript. The \mg{} can process multiple files in parallel and can
further parallelize the processing of each file. For example, image and catalog tiles
can be generated in parallel. See Figure \ref{fig:overview} for a graphical
representation of the \mg{}. The \mg{} performs two major tasks when generating
an interactive web website: parsing image files and parsing catalog files.
\fitsmap{} dynamically generates the \mv{} files using the image and
catalog data information.

\subsubsection{Parsing Image Data}
\label{sec:mg-image-parsing}

The first step in parsing the image data is to convert the image data into a
\texttt{numpy} \citep{harris2020} array. The \mg{} converts \texttt{FITS} files using
\texttt{astropy}
\citep{astropy2013,astropy2018} and converts traditional image files (\texttt{PNG},
\texttt{JPG},
etc.) using \texttt{pillow} \citep{clark2015}.
The \mg{} builds tiled representations, in
parallel if desired, of the array data at each zoom level in a structured set of
directories compatible with the \mv{}.

\subsubsection{Parsing Catalog Data}
\label{sec:mg-catalog-parsing}

The \mg{} can parse plain text files that are delimited (\texttt{CSV}, \texttt{TSV}, etc.)
and
where the first line contains the column names.  Catalog source
locations can be stored in either $[x, y]$ (image pixel coordinates) or $[\alpha,\delta]$
(right ascension and declination).
If the coordinates are given in $[\alpha,\delta]$,
then a \texttt{FITS} file containing a reference World Coordinate System (\texttt{WCS}) to
translate the sky coordinates to image coordinates must be provided.
After reading the catalog data, the \mg{}
clusters the data at every zoom level using a python port \citep{hausen2022a} of
the \texttt{supercluster} \citep{mapbox2022a} JavaScript library.
Then the \mg{} builds a tiled representation of the catalog
source and cluster locations at every zoom level. The tiled catalog source
locations are encoded and stored using the MapBox Vector Tile Format
\citep{mapbox2022b}, a compact binary representation of structured data used to
represent geometric objects in map tiles. Other source attributes (effective
radius, magnitude, etc.) are stored in a separate JavaScript Object Notation
(\texttt{JSON}) file for each source in a dedicated directory for each catalog.

\subsection{\mv{}}
\label{sec:mv}
\begin{figure}[h]
    \centering
    \includegraphics[width=\linewidth]{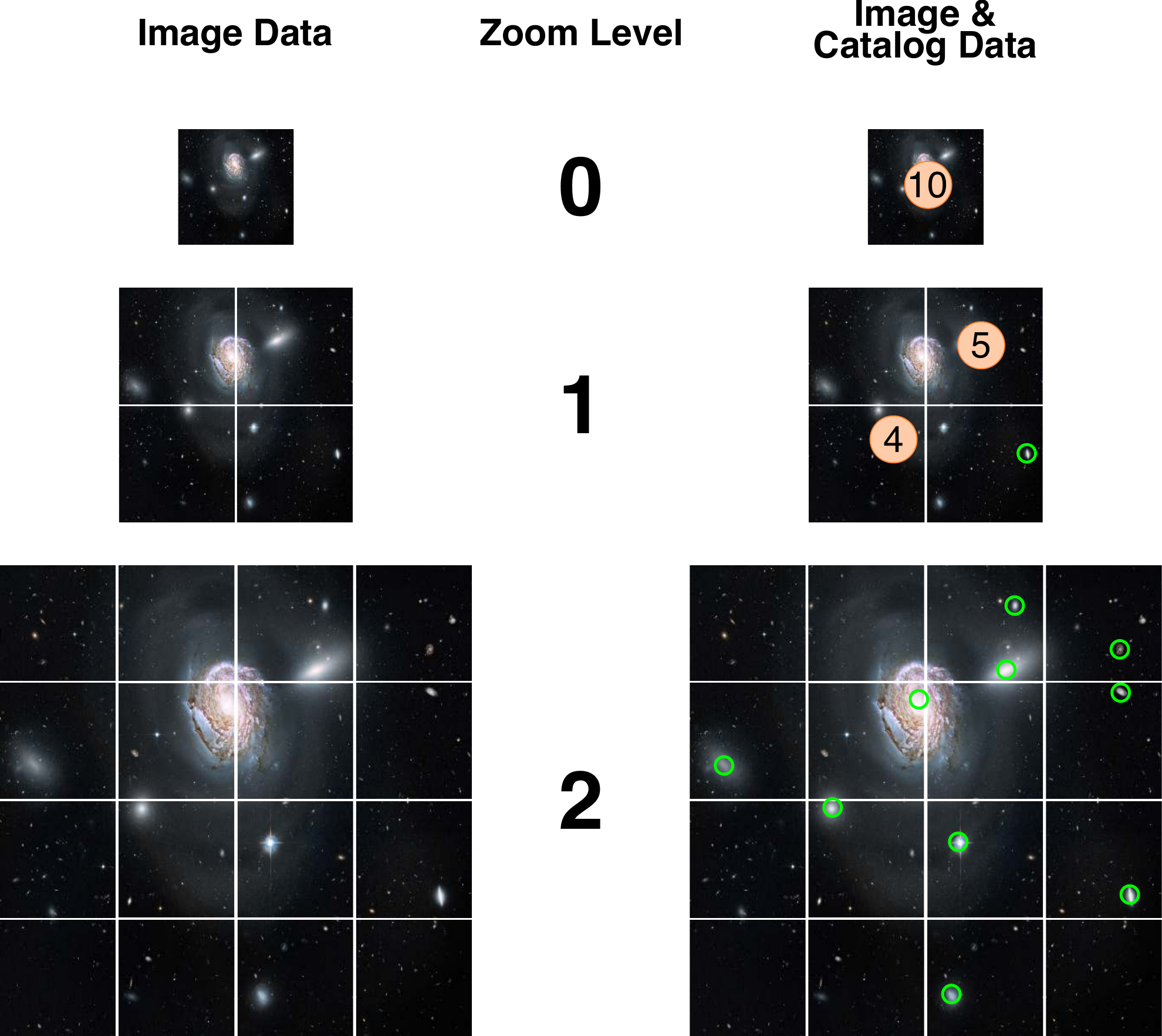}
    \caption{An illustration of tiling image and catalog data at three zoom
             levels (0, 1, 2). (left) Illustration of tiling image data only.
             (right) Illustration of tiling image and catalog data. For
             simplicity, not all sources in the image are labeled with a marker.
             \fitsmap{} computes the cluster assignments at every zoom level and
             stores them as markers in their respective tile. When the \mv{}
             requests the catalog data associated with a particular tile, it
             will retrieve either a source marker or a cluster marker and
             renders it appropriately. Source image credit: NASA.}
    \label{fig:example}
\end{figure}

\begin{figure}[h]
    \centering
    \includegraphics[width=\linewidth]{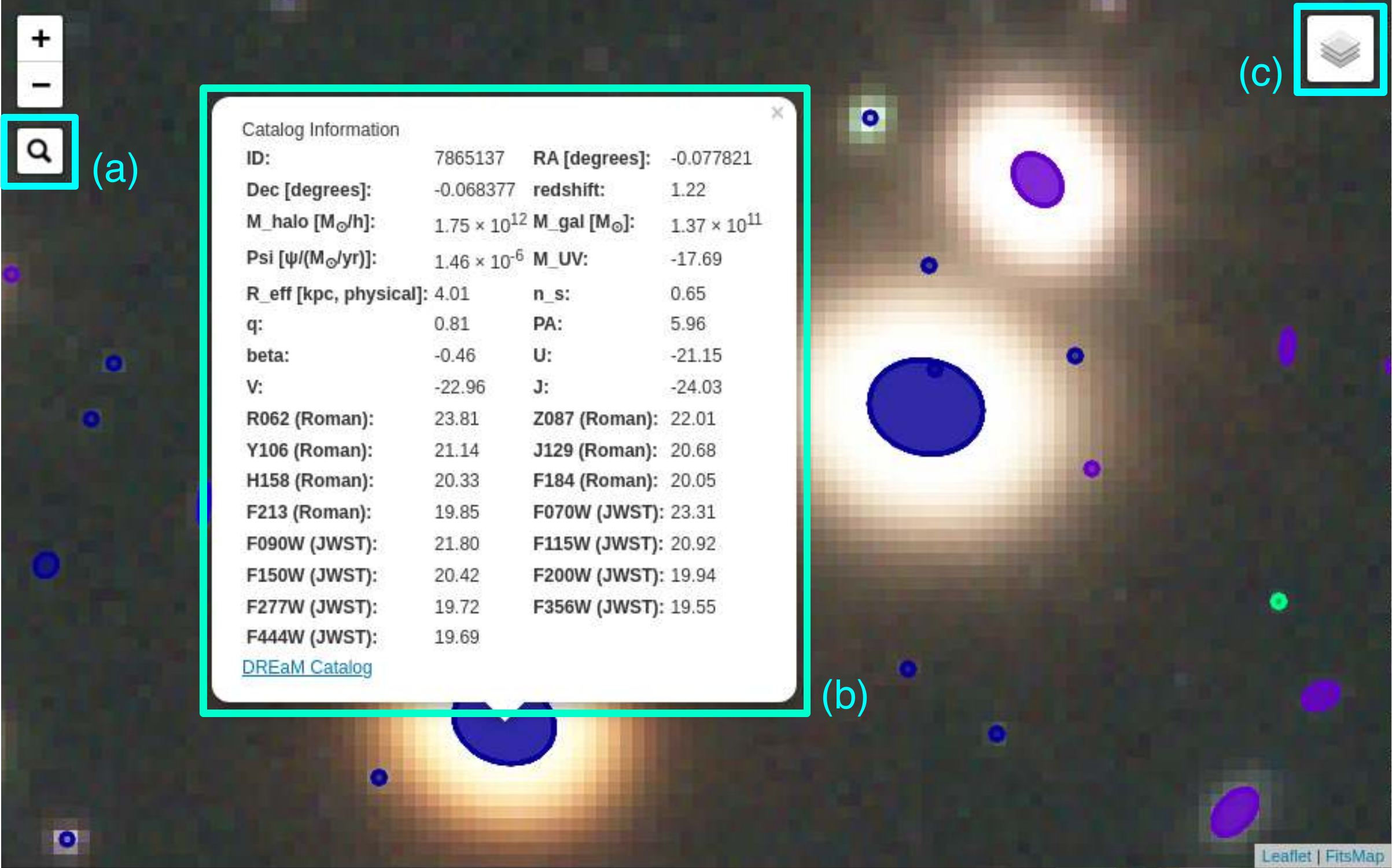}
    \caption{An example of the \fitsmap{} interface. (a) The search function
             button that searches catalogs by \texttt{id} (see Section
             \ref{sec:mv}). (b) A marker pop-up displaying catalog data
             associated with the indicated source. (c) The \texttt{Leaflet}
             layer control
             that allows users to switch between display images and catalog
             overlays. Image Credit: \cite{drakos2021}.}
    \label{fig:interface}
\end{figure}

\begin{sloppypar}
The \mv{} consists of the supporting \texttt{HTML},
\texttt{CSS}, image, and JavaScript files that
render the image and catalog data. The \mv{} is built using \texttt{Leaflet}
\citep{Agafonkin2022}. \texttt{Leaflet} is an open-source and flexible framework for
displaying large amounts of data. \texttt{Leaflet} supports rendering image tiles,
and to
support tiled marker data \fitsmap{} uses a custom-implemented \texttt{Leaflet} layer
called \texttt{L.GridLayer.TiledMarkers}, which is packaged with \fitsmap{}. The
\mv{} fetches and renders markers (sources and clusters) using the data
stored in each of the encoded tile files; see Figure \ref{fig:example} for an
illustration of image and catalog tiling. When a source marker is selected,
the \mv{} fetches additional information about the source from its corresponding
\texttt{JSON}
file (see Section \ref{sec:mg-catalog-parsing} and Figure \ref{fig:overview})
and renders it in a pop-up above the marker (see Figure \ref{fig:interface}).
The marker pop-ups support \texttt{HTML} to enable the
inclusion images and styling using a
custom \texttt{CSS}.
The \mv{} also includes a search function that can be used to query
any of the parsed catalogs by their \texttt{id} column.
The search functionality
is implemented using a custom search backend to the popular Leaflet Control
Search plugin \citep{cudini2022} that is packaged with \fitsmap{}. To search by
\texttt{id}, the \mv{} sends a \texttt{GET} request for a \texttt{JSON} file named
\texttt{\{id\}.json} in each catalog's extra \texttt{JSON} source file directory in
\texttt{catalog\_assets} (see Figure \ref{fig:example}). If the request returns
\texttt{404} (file not found), then the \texttt{id} does not exist. If the request
returns \texttt{200} (OK), then the desired object does exist
and the \mv{} can extract
location information stored in the file to pan and highlight the source.
\end{sloppypar}

\section{Performance}
\label{sec:performance}

As described in Section \ref{sec:methods}, to reduce the computational
requirements of both the server and the client \fitsmap{} tiles the marker data
such that the client only renders the markers that are currently in view
and unclustered at the current zoom level. In
this section, the performance of \fitsmap{} is evaluated in two ways. First, the
\mg{} is evaluated with respect to computational and storage requirements.
Second, the performance of the \mv{} is evaluated on both desktop and simulated
mobile environments using website profiling tools.

\subsection{\mg{} Performance}
\label{sec:mg-performance}

The \mg{} bears the largest computational burden within the \fitsmap{}
architecture. \fitsmap{} tiles an image by taking subsets of the input image
file and interpolates them into tile-sized images ($256^2$ pixels). \fitsmap{}
optimizes the tiling process in two ways. First, image tiling is parallelized
over tiles to provide a significant speedup.
Second, to ensure the entire input image does
not have to load into memory \fitsmap{} offers the option to limit
how far out the map may be zoomed.
\fitsmap{} processes catalog data in three steps:
parsing, clustering, and tiling (see Figure \ref{fig:overview}). Both parsing
and tiling are parallelizable in \fitsmap{}. The storage requirements of
clustering and tiling source catalog information are not easily predicted before
generating the map, but the storage requirements for the \texttt{JSON}
storage of catalog
sources with $r$ rows and $c$ columns
can be estimated per source using

\begin{equation}
    \label{eqn:json_cost}
    M_{j}(r, c) = s(r) + s(r_{xy}) + s(c) + n_{f} + 6n_{c} + 30
\end{equation}

\noindent
where
$s(\dot)$ is a function that returns the number of bytes required to encode the
argument as a string (excluding quotation marks, i.e. $s(3) = \texttt{sizeof}(``3") = 1$),
$r_{xy}$ is the location of the source represented by row $r$ in the image,
$n_f$ is the number of characters in the file name excluding the file extension,
$n_c$ is the number of columns in the catalog, and $30$ is the number of bytes
required to build the boilerplate data for the \fitsmap{} format. For example,
consider the catalog in Listing \ref{listing:catalog} below:

\begin{lstlisting}[label=listing:catalog,caption=An Example Catalog File,frame=tb]
example.csv
-----------
id,x,y,redshift
1,10,15,0.7
\end{lstlisting}

\noindent
The source in Listing \ref{listing:catalog} translates into the following inputs
to Equation \ref{eqn:json_cost}: $c=[\textrm{id,x,y,redshift}]$ represents the
column values and $r=[1,10,15,0.7]$ represents the row values. Each term
dependent on $r$ and $c$ in Equation \ref{eqn:json_cost} yields the following:
$s(r)=8$, $s(r_{xy})=4$, $s(c)=12$, $n_f=7$, $6n_c=24$, totaling $55+30=85$.
This indicates that the cost to store the source in \texttt{JSON} for \fitsmap{}
is $85$ bytes.

Calculating the total storage cost of tiling and clustering a catalog is
difficult \textit{a priori} as clustering depends on the spatial distribution of
sources in the image. Additionally, the Mapbox Vector Tile format is highly
optimized to account for data type and repeating values. However, the storage
requirements per tile containing $n_S$ individual sources and $n_C$ clusters of
sources can be estimated using

\begin{equation}
    \label{eqn:tile_cost}
    M_t(n_S, n_C) \approx 99 + 53n_S + 56n_C.
\end{equation}

\noindent
For example, a tile containing a single source and a single cluster will be
approximately $208$ bytes depending on the precision and repetition of catalog
values.

To provide an example of
the storage and computational requirements for using \fitsmap{}, a
sample map was made using a $50,000^2$-pixel \texttt{FITS} image ($10\textrm{GB}$) and a
sample catalog ($176.1\textrm{MB}$) containing $378,082$ sources with $57$
columns. Among five independent trials,
generating the map on an Intel Xeon E5-2698 system with
$232\textrm{GB}$ of RAM and processing the catalog and image in parallel with
each process using six workers took $213.3522\pm4.5727$ seconds. The total
data size of the input was approximately $10.2\textrm{GB}$, and the total of the
generated website was approximately $1.8\textrm{GB}$. The main reason for the
storage reduction is the conversion of \texttt{FITS} to \texttt{PNG}
for the image tiles. The
catalog data size grew from the input $176\textrm{MB}$ to approximately
$1.5\textrm{GB}$ upon output, and constituted most of the output data.

\subsection{\mv{} Performance}

The \mv{} is written entirely in JavaScript and only requires a web server to
render the output from the \mg{}. \mv{} performance is measured by rendering the
website generated in Section \ref{sec:mg-performance}, which contains
a $50,000^2$
pixel \texttt{FITS} image ($10\textrm{GB}$) and a sample catalog ($176.1\textrm{MB}$)
containing $378,082$ sources with $57$ columns. Performance is evaluated on a
locally hosted web server using the built-in Python \texttt{http.server}
package. \fitsmap{} is evaluated using the Google Chrome website auditing tool
\texttt{Lighthouse} to calculate the \textit{First Contentful Paint} (FCP) and
\textit{Time to Interactive} (TTI) for both desktop and simulated mobile
environments. \textit{FCP} measures the amount of time between the page request
and the first rendered element. \textit{TTI} measures the amount of time between
a page request and the page becoming interactive. Running \texttt{Lighthouse} in desktop
mode, we found
both FCP and TTI were 0.6 seconds. Running \texttt{Lighthouse} in mobile mode,
we found FCP
was 2.7 seconds and TTI was 2.9 seconds. Additionally, the Chrome performance
profiling tool was used to monitor memory consumption during a session
consisting of loading the page, displaying the catalog, and zooming in until
clustering is no longer applied and all local
source locations are rendered. In the trial
run for both mobile and desktop, the \mv{} never exceeded $16\textrm{MB}$
and stayed well within the RAM capacity of desktops and mobile devices.

\section{Conclusion and Future Work}
\label{sec:conclusion}
\begin{sloppypar} 
    This work introduced \fitsmap{}, a simple, lightweight tool for generating
    interactive web-based visualizations of astronomical images and their associated
    catalogs. \fitsmap{} uses a novel approach to serve catalog data that
    precomputes cluster information for the entire image at all zoom levels and
    displays the cluster and source information based on currently visible image
    tiles. Advantages of precomputing and tiling the catalog data include 1) the
    visualization only requires a simple web server and 2) the produced website
    loads quickly and proves responsive for large catalogs, even on mobile devices.
    The techniques developed in this work have already been used on public-facing
    websites. Both \citep{hausen2020} and \citep{drakos2021} leverage \fitsmap{} to
    generate interactive visualizations displaying their data products (See
    \url{https://purl.org/fitsmap/morpheus} and \url{https://purl.org/fitsmap/dream}
    respectively). Potential drawbacks to \fitsmap{} include the increased storage
    requirements above the input image and catalog data sizes that owe to the tiling
    method. The \fitsmap{} clustering and tiling requires preprocessing of the
    catalog data, but the preprocessing only needs to be performed once per website.
    The development of \fitsmap{} is active and on-going. Future work includes
    allowing for more complex shapes in catalog data such as polygons, allowing
    users to change \texttt{FITS} image scaling on the fly in the \mv{}, and
    minimizing storage usage.
\end{sloppypar}

\section{Acknowledgements}
\label{sec:acknowledgements}

We thank Daniel Eisenstein for suggesting the concept that eventually became
\fitsmap{}.
The authors acknowledge use of the lux supercomputer at UC Santa Cruz, funded by
NSF MRI grant AST 1828315, and support from NASA grant 80NSSC18K0563 and the
NIRCam science team.

\bibliography{refs}

\end{document}